\documentclass[%
 reprint,
superscriptaddress,
 amsmath,amssymb,
 aps,
longbibliography
]{revtex4-1}

\usepackage{graphicx}
\usepackage{dcolumn}
\usepackage{bm}
\usepackage{setspace}
\usepackage{xcolor}
\usepackage{soul}

\usepackage[english]{babel}

\begin{document}

\preprint{APS/123-QED}

\title{Cryogenic photonic resonator with $10^{-17}$/s drift}

\author{Wei Zhang}
	\email{Present address: Jet Propulsion Laboratory California Institute of Technology, 4800 Oak Grove Drive, Pasadena, California 91109-8099, USA}
	\affiliation{National Institute of Standards and Technology, 325 Broadway, Boulder, Colorado 80305, USA \\}
\author{William R. Milner}
	\affiliation{JILA, National Institute of Standards and Technology and University of Colorado, Boulder, CO, USA \\}
\author{Jun Ye}
    \affiliation{JILA, National Institute of Standards and Technology and University of Colorado, Boulder, CO, USA \\}
\author{Scott B. Papp}
    \email{scott.papp@nist.gov}
    \affiliation{National Institute of Standards and Technology, 325 Broadway, Boulder, Colorado 80305, USA \\}
    \affiliation{Department of Physics, University of Colorado, Boulder, Colorado 80309, USA\\}

\date{\today}

\begin{abstract}
Thermal noise is the predominant instability in the provision of ultrastable laser frequency, \textcolor{black}{referencing to an optical cavity}. Reducing the thermal-noise limit of a cavity means either making it larger to spread thermal fluctuations, reducing the sensitivity of the cavity to temperature, or lowering the temperature. We report on a compact photonic resonator made of solid fused silica that we cool in a cryogenic environment. We explore a null in the resonator’s frequency sensitivity due to the balance of thermal expansion and thermo-optic coefficients at a temperature of 9.5 K, enabling laser stabilization with a long-term frequency drift of 4 mHz/s on the 195 THz carrier. The robustness of fused silica to cryogenics, the capability for photonic design to mitigate thermal noise and drift, and operation at a modest 9.5 K temperature offer unique options for ultrastable laser systems. 

\end{abstract}

\pacs{Valid PACS appear here}
\maketitle



Frequency-stabilized lasers based on optical resonators that are engineered for low sensitivity to environmental fluctuations exhibit high coherence, signified by a combination of narrow integrated linewidth of the laser spectrum, low phase noise of laser frequency fluctuations, and low laser frequency drift. The highest performance cavity-stabilized lasers, reaching a fractional frequency stability (FFS) of 10$^{-16}$ at 1 s and frequency drift of 100 $\mu$Hz/s, are critical for optical atomic clocks and the future potential to redefine the SI second \textcolor{black}{~\cite{Bloom2014,RIEHLE2015506,Beloy2021, Milner_PRL_2019}}. Compact and manufacturable stabilized lasers are also useful for applications like data communication~\cite{Brodnik2021}. Such cavities or resonators take different forms from traditional Fabry-Perot (FP) cavities with an evacuated gap between the mirrors ~\cite{Matei2017,ZhangPRL2017} to bulk resonators, including photonic resonators ~\cite{ZhangLPR2020}, microrod ~\cite{Zhang_PRAppleid_2019}, whispering-gallery mode resonators~\cite{AlnisPRA2011,LiangNC2015}, on-chip microresonators with waveguides~\cite{LiuOptica2022}. These resonator technologies, subject to an essentially common set of physical considerations, enable a range of laser-stabilization performance, potential for photonic integration, and design for application tradeoffs. 

At the center of engineering these systems is reduction of their sensitivity to the environment, including temperature-induced material expansion and change in refractive index, vibration-induced resonator deformation, and fluctuations from air pressure. Fabry-Perot (FP) systems made from discrete parts held in vacuum offer a rich optimization palette, and indeed exceptionally low vibration sensitivity ~\cite{ChenPhysRevA2006}, low thermal expansion~\cite{Matei2017,ZhangPRL2017}, and feed-forward control techniques~\cite{LeibrandtOE2011} have been realized. For example, the cryogenic silicon cavity attains vibration sensitivity of $10^{-11}$/g and drift on 10 $\mu$Hz/s level, where g is gravity ~\cite{Robinson19Optica}. An important enabling feature for optimization of FP systems is the tight tolerance between the geometry of the mirrors and the transverse mode pattern. With photonic resonators and waveguide microresonators that also feature transverse mode control from geometry, the substantial temperature coefficient of refractive index is the most important challenge for attaining laser-frequency stability. Indeed, even the best current photonic devices do not offer stability much better than 10$^{-12}$ at 1 s despite experiments with complex environmental controls~\cite{AlnisPRA2011}, athermal resonators, optimized material properties, and feed-forward controls~\cite{Lim2019,ZhaoOptica2021}

In the use of resonators for laser stabilization a few considerations predominantly determine system design, including low intrinsic optical loss to support narrow resonator mode linewidth for frequency discrimination ~\cite{Wong85} and the thermal-noise limit of resonator mode fluctuations ~\cite{Numata2004, Matsko2007}. Interestingly, both the traditional FP mirrors based on ion-beam sputtered tantalum pentoxide coatings and more recent photonic-integrated devices support sufficiently low loss~\cite{Zhang_PRAppleid_2019, ZhangLPR2020,SternOL2020} for discrimination below the 1 Hz$^2$/Hz level of frequency noise power spectral density. Furthermore, laser stabilization with both traditional FP and photonic resonators reach their fundamental thermal-noise limits, consistent with particular geometry and material properties. For example, the cryogenic silicon cavity and our room temperature photonic resonators reach a thermal noise \textcolor{black}{limited linewidth} of 10 mHz ~\cite{Matei2017} and 18 Hz ~\cite{Zhang_2023}, respectively. This highlights a tradespace in system design to comprehensively explore in which cryogenic operation, system size and power consumption, and material properties must be jointly understood. Current experiments are also searching for materials with lower thermal noise contribution, for example metal-oxide mixtures~\cite{PhysRevD.105.102008}

\begin{figure}[ht]
\centering
\includegraphics[width=\linewidth]{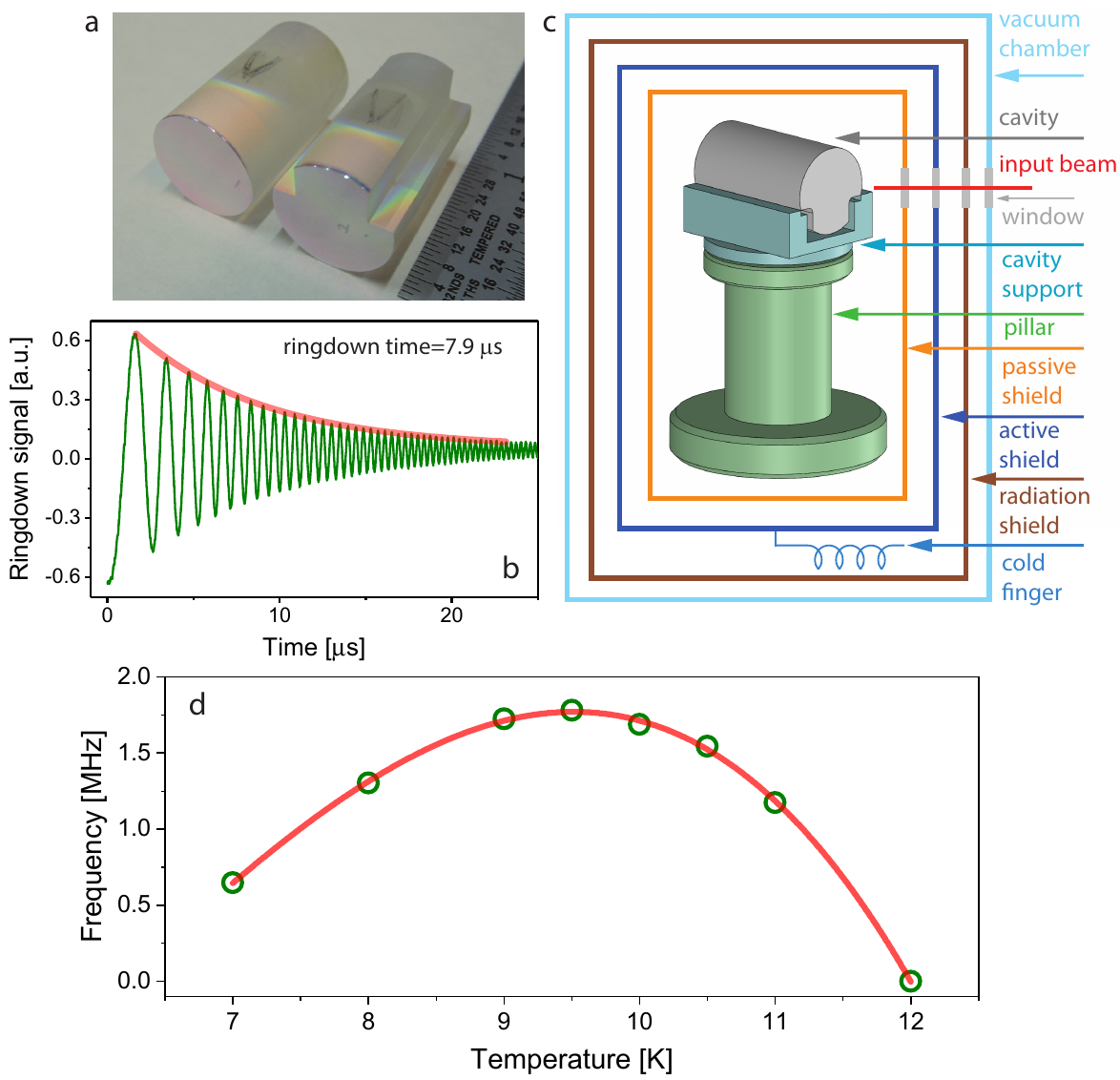}
\caption{(a) Photo of photonic resonators, including with mushroom cross section. (b) Ringdown measurement of resonator with 7.9 $\mu$s corresponds to finesse of 100,000. (c) Schematic of the cryocooler and thermal shields. (d) Frequency variation as function of the cryocooler temperature. \textcolor{black}{Green dot, amplitude of the photonic resonator frequency change in exponential fitting when the a step temperature change is applied on active shield. Red curve, the polynomial fitting.}}
\vspace{0pt}
\label{fig1}
\end{figure}

Here, we explore laser-frequency stabilization to a photonic resonator at cryogenic temperature, which offers the advantages of reduced thermal noise and a null of temperature sensitivity, due to the balance of thermal expansion and thermo-optic coefficients.
Moreover, the photonic resonator is well adaptable to the cryogenic environment due to its small size, robust construction, and internal containment of the optical mode. We use a photonic resonator composed of a bulk, fused-silica substrate with superpolished and high-reflection-coated endfaces to support resonator modes. The resonator supports a finesse of 100,000 and a temperature sensitivity of only $2.3 \times 10^{-9}$/$K^2$ at an absolute temperature of 9.5 K. Operating the photonic resonator in this cryogenic environment, we achieve laser-frequency stabilization with a low absolute drift rate of 4 mHz/s or fractionally $10^{-17}$/s, measured over a 9 day period by way of comparison to the cryogenic silicon cavity at JILA ~\cite{Matei2017} over a stabilized fiber link~\cite{Beloy2021}. Our experiments highlight a new tradespace in cavity stabilized lasers in which photonic design accesses nearly record performance with modest cryogenic requirements.

Since the optical mode and material properties of our photonic resonator are well characterized, we can predict the laser-frequency stability that may be achievable in our system. Indeed, operating at the 9.5 K temperature, the thermal noise is contributed by the thermorefractive noise of the resonator and the Brownian of the coating, resulting in $1 \times 10^{-15}$ at $\tau$=0.1 s, $6 \times 10^{-16}$ at $\tau$=1 s, and $4 \times 10^{-16}$ for $\tau>$ 10 s, where $\tau$ is the averaging time.

Figure 1 presents the photonic resonator and its high finesse, our cryogenic system, and measurement of the thermal sensitivity null. The photonic resonator (Fig. 1a) is a 25.4 mm long, 15 mm diameter, fused-silica cylinder, which is constructed from Suprasil 3001. Here, the material properties must satisfy a few aims, particularly low optical absorption from the material and impurities, low optical scattering from inclusions and spatially dependent index of refraction variation, and the capability to support superpolishing and low-loss mirror film deposition.  The endfaces of the cylinder are superpolished to a plano and convex shape (radius of curvature 0.5 meter), respectively, and we apply low-loss, high-reflective, dual band optical coatings designed for 1550 nm and 780 nm. These coatings are characterized by optical field ringdown measurement. We have developed fabless manufacturing of the devices in batches of ten or more by engaging with vendors for the fused silica substrates, machining and polishing the substrates, and design and deposition of the mirror coatings. 

We place the resonator in a commercial cryostat with operation from 3.5 K to room temperature, and we stabilize a laser to the resonator by use of a Pound-Drever-Hall setup. We use external cavity, continous-wave laser (RIO Planex) either at 1550 nm or 1542 nm, since its narrow linewidth and frequency controls are well-matched to the photonic-resonator modes. We ensure the laser is well isolated and linearly polarized in polarization maintaining fiber. We phase modulate the laser and reduce residual amplitude modulation in fiber before launching a free-space beam that we mode match and align to the resonator. We obtain high contrast transmission and reflection outputs from the photonic resonator that we photodetect with low-noise avalanche photodiodes. By rapidly modulating the input laser frequency, we monitor the cavity ringdown with 7.9 $\mu$s in the reflection field; see Fig. 1b. The photonic resonator supports a mode finesse of 100,000 in operation from 3.5 K to 300 K, which we observe in the cooldown process of the cryostat. 

Previous experiments with photonic resonators ~\cite{ZhangLPR2020} have demonstrated their thermal-noise-limited performance. However, at room temperature, fused silica exhibits a $10^{-5}$/K refractive index variation, therefore the resonator’s mode frequencies are exceptionally sensitive to temperature fluctuations and drift. Moreover, the thermal expansion coefficient of fused silica, approximately $0.5\times10^{-6}$/K, can largely be ignored. The work of Ref.~\cite{PhysRevA.80.021803} indicated a null in the mode temperature sensitivity of a fused-silica microresonator around 10 K. The null is obtained due to a balance of the temperature variation of refractive index, which is slightly negative at 9.5 K, and the positive thermal expansion.

We use a two-stage Gifford-McMahon closed-cycle cryocooler with customization for temperature stability. Figure 1c shows a diagram of the cryocooler configuration. Since the temperature fluctuations of the cryocooler are typically 20 mK at an averaging time of 6 s, we apply a custom-designed thermal shield between the cold finger and the resonator to suppress temperature fluctuations on the resonator. The first stage of the cryocooler conducts the radiation shield (30 K, Aluminum), reducing the radiation between the vacuum chamber (300 K, Aluminum) and the rest in the cryostat. The second stage of the cryocooler is thermally coupled to the bottom of the first layer of the thermal shield (copper), labeled as active shield, through the cold finger, which is a mechanically flexible copper braid. The temperature of the active shield is controlled and stabilized at the milliKelvin level by use of a temperature servo in which a resistive heater serves as the thermal actuator. The passive shield (copper) is built in the active shield, thermally isolated by a short pillar ($\sim$1 mm in height) made of holmium copper (HoCu$_2$, not shown in Fig. 1c), to increase the thermal damping. The resonator is installed in a U-shaped support (made of PEEK) with four small supporting pads of rectangular shape, machined in the PEEK support. The PEEK support sits on a pillar (made of stainless steel) in the passive shield. 

To explore the temperature sensitivity null of the photonic resonator, we monitor the resonant frequency of a mode as we vary the temperature of the device. We record the heterodyne beatnote between a 1550 nm CW laser that is locked to the photonic resonator and a FP stabilized reference laser with $\sim 0.2$ Hz/s drift. We apply a step change on the temperature of the active shield, which results in a slow equilibration of the photonic-resonator temperature with a time constant of approximately 10,000 s, \textcolor{black}{measured by the corresponding frequency change with exponential fitting.} This relatively long time constant of the thermal shield to the actively stabilized cryocooler base temperature indicates the high thermal isolation of our system design. To explore the null of temperature sensitivity of the resonator, the frequency change on the beatnote as the function of temperature of active shield is recorded; Fig. 1d presents the measurement from 7 K to 12 K. \textcolor{black}{Each data point (green) represents the amplitude of the exponential fitting. This measurement with the polynomial fitting (red) shows a tuning temperature around 9.5 K} where the first order of frequency change is null due to the balance between the resonator thermal expansion and the refractive index. At 9.5 K, the second-order sensitivity is as low as $2.3 \times 10^{-9}$/$K^2$, which is comparable with typical ULE performance at room temperature. 



\begin{figure}[h]
\centering
\includegraphics[width=\linewidth]{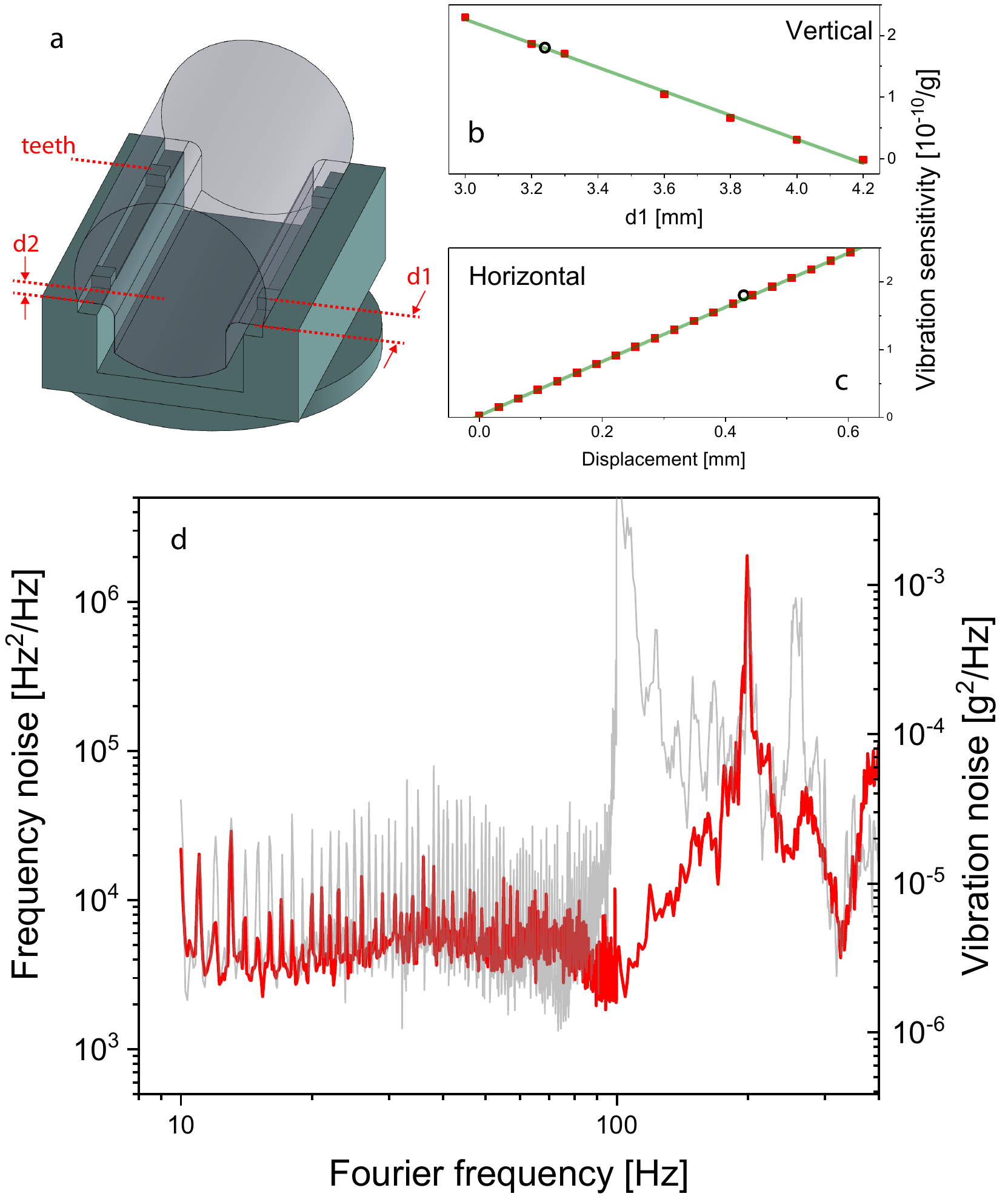}
\caption{(a) Mechanical schematic of the photonic resonator with vibration insensitivity design and its mounting structure. (b) The vertical vibration sensitivity as function of d1 \textcolor{black}{(red-closed circles) and the linear fitting (green) shows a slope of $2 \times 10^{-10}$/g/mm.}  The measured sensitivity corresponds (black-open circle) to d1=3.22 mm. (c) The horizontal vibration sensitivity as function of the displacement of the beam from resonator axis \textcolor{black}{(red-closed square) and the linear fitting (green) shows a slope of $4 \times 10^{-10}$/g/mm.} The measured sensitivity (black-open circle) indicates the displacement is 0.45 mm (d) Frequency noise of the beatnote between the silicon cavity
and the resonator (red), and the vibration-induced
frequency noise (gray) calculated from the measured
vibration noise of the cryocooler environment.}
\label{fig3} \vspace{-9pt}
\end{figure}

Because vibration from the cryocooler causes frequency noise of the photonic resonator modes, we design the resonator shape and its mounting structure made of PEEK material to minimize its vibration sensitivity; see Fig. 2 for details and measurements of our vibration isolation design. For the resonator, the axial cross section has a mushroom shape by machining two ridges in the lower half of the cylinder. Based on finite-element analysis (FEA), we adjust the contact position, i.e. the distance from the cavity to the surface of the support (labeled as d1 in Fig. 2a), and ridge dimension, i.e. the offset of cut-in in the mushroom shape (d2 in Fig. 2a), to optimize the vibration sensitivity $k_v$ along the vertical direction and $k_h$ in the horizontal plane. According to the FEA simulations, we choose d1=4.2 mm and d2= 2 mm.

To measure the vibration sensitivity of the resonator in the PEEK support, we assemble these two pieces (Fig. 2a) and place them on an active vibration isolation (AVI) table. We apply a modulation signal with vector signal analyzer to drive the AVI table, and we calibrate the motion of the AVI with an accelerometer. The corresponding photonic resonator frequency modulation is measured by the heterodyne beatnote between the laser locked on the photonic resonator and the FP stabilized reference laser. The $k_v$ parameter is measured to be $1.8 \times 10^{-10}$/g (g=9.8~m/s$^2$), and we get the same value for $k_h$. To interpret the difference between the FEA simulation and measurement, Fig. 2b presents $k_v$ as function of d1 (here we assume d2=2 mm), with a slope of $2 \times 10^{-10}$/g/mm. The measured $k_v$ corresponds to d1 with a 0.9 mm offset from the design. Figure 2c shows $k_h$ as function of the transverse displacement of the beam from the cavity axis (here we assume d1=4.2 mm and d2=2 mm), with a slope of $4 \times 10^{-10}$/g/mm. The actual beam position might be 0.45 mm offset from cavity geometry center. By comparing the simulation and measurement, either $k_v$ or $k_h$ offers a reasonable margin, though we would explore more geometry and support strategies in the future to reduce the vibration sensitivity. 

Upon assembling our system, operating the cryocooler to access the 9.5 K null temperature, frequency stabilizing the laser to the resonator, and characterizing the laser with respect to the reference laser, we find that cryocooler vibrations pervasively dominate the spectrum of noise on the stabilized laser. In this system, we employ the AVI table in active cancellation mode, directly beneath the thermal-shield assembly. Here, we demodulate the heterodyne beatnote, record the time-domain waveform, and calculate the frequency-noise spectrum via a fast Fourier transform (FFT); see the red line in Fig. 2d. We also compare the predicted frequency-noise spectrum (gray line in Fig. 2d) based on the measured vibration sensitivity of the photonic resonator and the measured vibration environment of the operating cryocooler. Though the vibration sensitivities have been optimized, it is obvious that the laser frequency noise is limited by the vibration induced by the cryocooler. 

\begin{figure}[h!]
\centering
\includegraphics[width=\linewidth]{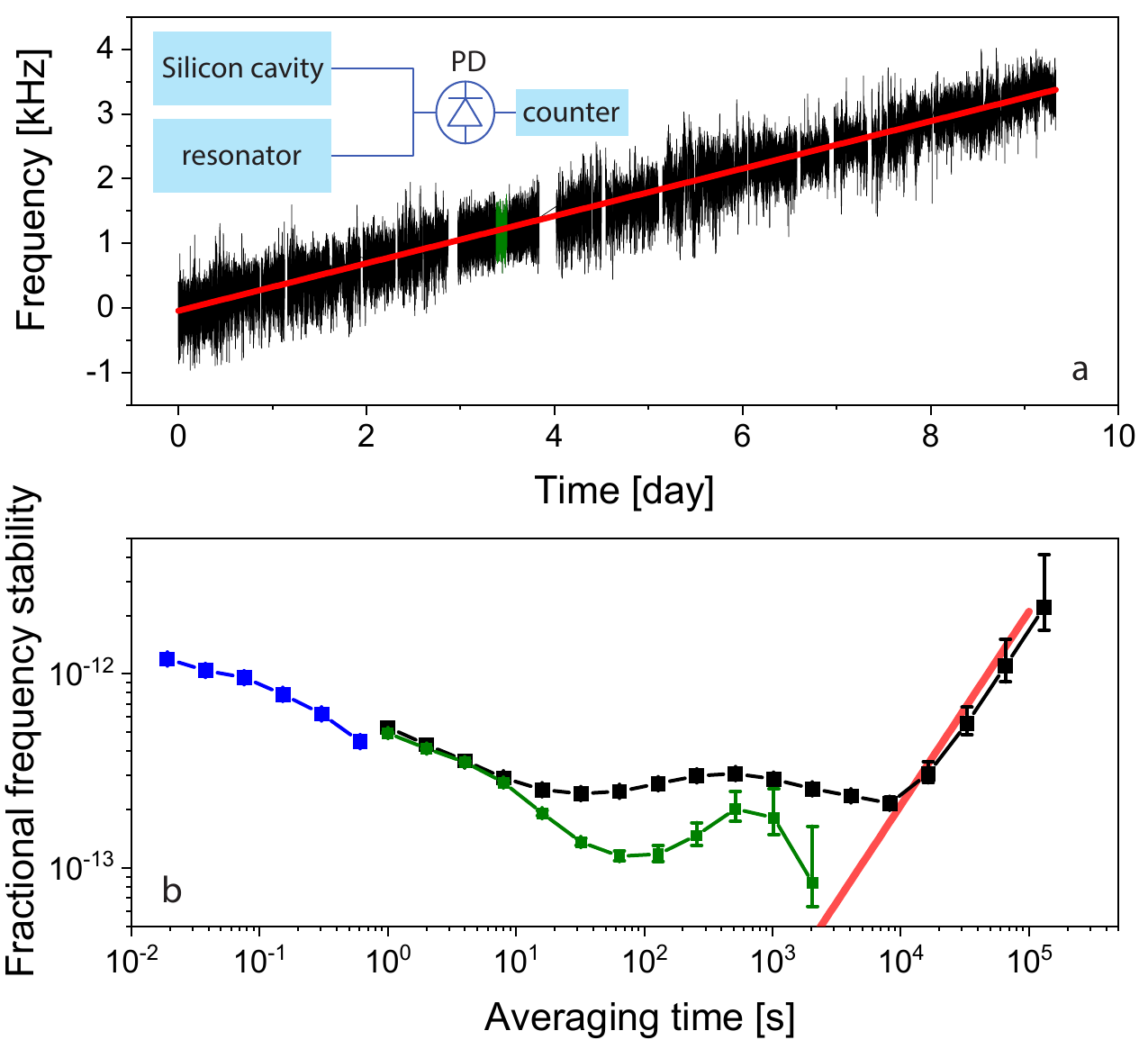}
\caption{(a) The 9 day record of the beatnote between the photonic resonator and the silicon cavity (black) and the linear fit at 4 mHz/s (red). A data section (10 000s, green) is selected to calcuate the frequency stability without obvious perturbation due to vibration \textcolor{black}{(b) Fractional frequency stability of the photonic resonator is derived from the beatnote measurement. Blue, 160 s data with 20 ms gate time; Black, 9 day data with 1 s gate time; Green, selected 10 000 s data in figure 3a with 1 s gate time. Red: the 4 mHz/s linear drift is dominant for averaging time above 10 000 seconds.}}
\label{fig2}  \vspace{-9pt}
\end{figure}

Though the short-term frequency noise is limited by vibration, we are still able to measure the long-term low drift rate of the laser referenced to the photonic resonator thanks to the null of temperature sensitivity and well-established thermal filter. We stabilize the frequency of a 1542 nm external cavity diode laser to the cryogenic photonic resonator. \textcolor{black}{As shown in the inset of Fig. 3a,} to evaluate the drift of the stabilized laser, we measure a heterodyne beatnote with a reference laser, which is generated at JILA with an evacuated silicon cavity ~\cite{Matei2017} and transmitted to NIST by a stabilized fiber link~\cite{Beloy2021}. The reference laser drift is $<$100 $\mu$Hz/s. The beatnote is recorded by a dead-time free frequency counter.  The black line in Fig. 3a presents a 9-day record of the heterodyne beatnote, using 1 s gate time of the frequency counter. The red line is the linear fitting of the beatnote, representing frequency drift of the photonic resonator mode. According to the fitting, the drift rate is consistently at 4 mHz/s, corresponding to a FFS of $2 \times 10^{-17}$/s. The FFS of the photonic resonator is presented by the \textcolor{black}{modified} Allan deviation of the beatnote. In Fig. 3b, the blue line is the short-term FFS computed by using 160 s data with 20 ms gate time, and the \textcolor{black}{black} line is for middle- and long-term FFS by using the 9-day data with 1 s gate time. The short-term FFS is dominated by the vibration from cryocooler, which is consistent with the measurement in Fig. 2d. The 9-day continuous measurement shows vibration-induced frequency spikes of $\approx 2$ kHz peak-to-peak.
As a result, the FFS is $3 \times 10^{-13}$ for at intermediate measurement time of 1 s$<\tau<10^4$ s. If we pick a certain time period, e.g. the 3-hour section highlighted in Fig. 3a in which the frequency spikes are slightly reduced, the corresponding FFS is $1 \times 10^{-13}$ as shown in green curve in Fig. 3b. The red line corresponds to the 4 mHz/s drift rate, which overlaps with the long-term FFS for $\tau>10^4$ s. 

In summary, we have explored laser stabilization with a photonic resonator \textcolor{black}{made of fused silica and} cooled to 9.5 K. The null of temperature sensitivity from photonic design of the resonator and the low absolute temperature makes possible a low fractional frequency drift of $2 \times 10^{-17}$/s and a thermal-noise limit of $4 \times 10^{-16}$ for $\tau>$ 10 s.
In this concept experiment, the cryocooler capacity was dramatically oversized in a commercially available and adaptable cryo system. An optimized cryo system could access a unique combination of system size and stabilized laser performance.


\begin{acknowledgments}

This work is a contribution of NIST and not subject to US copyright. This research has been funded by AFOSR FA9550-20-1-0004 Project Number 19RT1019, NSF Quantum Leap Challenge Institute Award OMA - 2016244, and NIST on a Chip. Mention of specific companies or trade names is for scientific communication only and does not constitute an endorsement by NIST.

\end{acknowledgments}

\bibliography{reference2.bib} 

\end{document}